\documentclass[twocolumn,iop]{openjournal}

\usepackage[british]{babel}

\usepackage{mathtools}
\usepackage{natbib}
\usepackage[colorlinks,allcolors=blue]{hyperref}

\begin{document}

\journalinfo{The Open Journal of Astrophysics}
\submitted{submitted August 2020}

\shorttitle{X-ray \emph{eROSITA} virtual sky}
\shortauthors{J. Comparat et al.}

\title{Full-sky photon simulation of clusters and active galactic nuclei in the soft X-rays for \emph{eROSITA}}

\author{Johan Comparat$^{\star1}$    }
\author{Dominique Eckert$^{2}$           }
\author{Alexis Finoguenov$^{3}$       }
\author{Robert Schmidt$^{4}$          }
\author{Jeremy Sanders$^{1}$          }
\author{Daisuke Nagai$^{5}$            }
\author{Erwin T. Lau$^{6}$              }
\author{Florian Kaefer$^{1}$           }
\author{Florian Pacaud$^{7}$           }
\author{Nicolas Clerc$^{8}$            }
\author{Thomas H. Reiprich$^{7}$         }
\author{Esra Bulbul$^{1}$           }
\author{Jacob Ider Chitham$^{1}$     }
\author{Chia-Hsun Chuang$^{9}$         }
\author{Vittorio Ghirardini$^{1}$       }
\author{Violeta Gonzalez-Perez$^{10}$  }
\author{Ghassem Gozaliasl$^{3}$        }
\author{Charles C. Kirkpatrick$^{3}$      }
\author{Anatoly Klypin$^{11}$          }
\author{Andrea Merloni$^{1}$          }
\author{Kirpal Nandra$^{1}$           }
\author{Teng Liu$^{1}$              }
\author{Francisco Prada$^{14}$           }
\author{Miriam E. Ramos-Ceja$^{1}$           }
\author{Mara Salvato$^{1}$          }
\author{Riccardo Seppi$^{1}$            }
\author{Elmo Tempel$^{17}$          }
\author{Gustavo Yepes$^{18}$           }

\affiliation{$^{1}$Max-Planck Institut f{\"u}r extraterrestrische Physik, Postfach 1312, D-85741 Garching bei M{\"u}nchen, Germany}
\affiliation{$^{2}$The Astronomy Department, University of Geneva, Ch. d’Ecogia 16, CH-1290 Versoix, Switzerland}
\affiliation{$^{3}$Department of Physics, University of Helsinki, Gustaf H\"allstr\"omin katu 2a, FI-00014 Helsinki, Finland}
\affiliation{$^{4}$Astronomisches Rechen-Institut, Zentrum f{\"u}r Astronomie der Universit{\"a}t Heidelberg, M{\"o}nchhofstrasse 12-14, 69120 Heidelberg, Germany}
\affiliation{$^{5}$Department of Physics, Yale University, New Haven, CT 06520, U.S.A.}
\affiliation{$^{6}$Department of Physics, University of Miami, Coral Gables, FL 33124, U.S.A.}
\affiliation{$^{7}$Argelander-Institut f{\"u}r Astronomie (AIfA), Universit{\"a}t Bonn, Auf dem H{\"u}gel 71, 53121 Bonn, Germany}
\affiliation{$^{8}$IRAP, Universit\'e de Toulouse, CNRS, UPS, CNES, Toulouse, France}
\affiliation{$^{9}$Kavli Institute for Particle Astrophysics and Cosmology, Stanford University, 452 Lomita Mall, Stanford, CA 94305, USA}
\affiliation{$^{10}$Astrophysics Research Institute, Liverpool John Moores University, 146 Brownlow Hill, Liverpool L3 5RF, UK}
\affiliation{$^{11}$Astronomy Department, New Mexico State University, Las Cruces, NM, USA}
\affiliation{$^{12}$Laboratoire d'Astrophysique, Ecole Polytechnique Fédérale de Lausanne (EPFL), Observatoire de Sauverny, CH-1290 Versoix, Switzerland}
\affiliation{$^{13}$Institute for Computational Cosmology and Centre for Extragalactic Astronomy, Department of Physics, Durham University, South Road, Durham, DH1 3LE, UK}
\affiliation{$^{14}$ Instituto de Astrof\'{\i}sica de Andaluc\'{\i}a (CSIC), Glorieta de la Astronom\'{\i}a, E-18080 Granada, Spain }
\affiliation{$^{15}$Univ Lyon, Univ Lyon1, Ens de Lyon, CNRS, Centre de Recherche Astrophysique de Lyon UMR5574, F-69230 Saint-Genis-Laval, France}
\affiliation{$^{16}$IRFU, CEA, Universit\'{e} Paris-Saclay, F-91191 Gif-sur-Yvette, France}
\affiliation{$^{17}$Tartu Observatory, University of Tartu, Observatooriumi 1, 61602 T\~oravere, Estonia}
\affiliation{$^{18}$ Departamento de F\'{\i}sica Te\'orica and CIAFF, Universidad Aut\'onoma de Madrid, 28049 Madrid, Spain}

\thanks{$^\star$ E-mail: \nolinkurl{comparat@mpe.mpg.de}}

\begin{abstract}
% context
The \emph{eROSITA} X-ray telescope on board the Spectrum-Roentgen-Gamma (SRG) mission will measure the position and properties of about 100\,000 clusters of galaxies and 3 million active galactic nuclei over the full sky.
%aim
To study the statistical properties of this ongoing survey, it is key to estimate the selection function accurately. 

% method
We create a set of full sky light-cones using the MultiDark and UNIT dark matter only N-body simulations. 
We present a novel method to predict the X-ray emission of galaxy clusters. 
Given a set of dark matter halo properties (mass, redshift, ellipticity, offset parameter), we construct an X-ray emissivity profile and image for each halo in the light-cone. 
We follow the \emph{eROSITA} scanning strategy to produce a list of X-ray photons on the full sky.

% result
We predict scaling relations for the model clusters, which are in good agreement with the literature. 
The predicted number density of clusters as a function of flux also agrees with previous measurements. 
Finally, we obtain a scatter of 0.21 (0.07, 0.25) for the X-ray luminosity -- mass (temperature -- mass, luminosity -- temperature) model scaling relations.

% perspective
We provide catalogues with the model photons emitted by clusters and active galactic nuclei. These catalogues will aid the \emph{eROSITA} end to end simulation flow analysis and in particular the source detection process and cataloguing methods.

\end{abstract}


\maketitle

%%%%%%%%%%%%%%%%% BODY OF PAPER %%%%%%%%%%%%%%%%%%
\section{Introduction}

As the most massive gravitationally bound objects in the Universe, Galaxy clusters have been widely used as probes for cosmology due to the exponential dependence of their abundance on cosmological parameters \citep[see][for reviews]{Voit2005RvMP...77..207V, Allen2011ARA&A..49..409A, Weinberg2013PhR...530...87W}. 

The \emph{eROSITA} X-ray telescope on board the Spectrum-Roentgen-Gamma (SRG) mission will measure the position and properties of about 100\,000 clusters of galaxies and 3 million active galactic nuclei over the full sky \citep{arxiv12093114_Merloni,2016SPIE.9905E..1KP}. 
The large \emph{eROSITA} cluster sample will improve cosmological constraints significantly \citep{Pillepich2018MNRAS.481..613P}. The improvement is contingent on our ability to constrain the X-ray emitting intracluster medium (ICM) profiles out large radii \citep{Walker2019SSRv..215....7W} that affect the mass measurements of galaxy clusters \citep{Pratt2019SSRv..215...25P}. 

Over the last decade, there have been numerous advances in modeling of the ICM properties and their X-ray emissions. 
In particular, cosmological hydrodynamical simulations have been the keys for understanding the roles of cluster astrophysics \citep[e.g.][]{Springel2001MNRAS.328..726S, Nagai2007ApJ...668....1N, Brun2014MNRAS.441.1270L, Dolag2016MNRAS.463.1797D, Dubois2016MNRAS.463.3948D, McCarthy2017MNRAS.465.2936M, Barnes2017MNRAS.465..213B, Barnes2017MNRAS.471.1088B, Cui2018MNRAS.480.2898C, Pillepich2018MNRAS.475..648P, Henden2018MNRAS.479.5385H} and their impacts on the X-ray scaling relations \citep[e.g.][]{Kravtsov2006ApJ...650..128K, Planelles2014MNRAS.438..195P}, cool-core properties \citep[e.g.][]{Rasia2015ApJ...813L..17R, Barnes2018MNRAS.481.1809B}, cluster morphology \citep[e.g.][]{Green2019ApJ...884...33G,Cao2020arXiv200610752C}, hydrostatic mass bias \citep[e.g.][]{Rasia2006MNRAS.369.2013R, Nagai2007ApJ...655...98N, Barnes2020arXiv200111508B}, AGN contamination of X-ray cluster signals \citep{biffi_2018MNRAS.tmp.2317B} using detailed mock X-ray simulations. 
However, because they are computationally expensive to run, they are less adapted to rapid iterations on the model towards full-sky \emph{eROSITA} predictions.

On the other hand, phenomenological and semi-analytical models \citep[e.g.][]{Shaw2010ApJ...725.1452S,BalagueraAntolnez2012MNRAS.425.2244B,Zandanel2014MNRAS.438..116Z,Flender2017ApJ...837..124F,Zandanel2018MNRAS.480..987Z,Clerc2018AA...617A..92C} have emerged as a complementary modeling approach to direct hydrodynamical simulations, 
allowing us to explore ICM physics in a relatively computationally efficient way. 
These models also allow us to generate full-sky photon-level cluster simulations that are essential for \emph{eROSITA} \citep{Zandanel2018MNRAS.480..987Z}.

In this paper, we present a novel  method to model X-ray cluster profiles. We adopt an empirical method where we sample ICM profiles from high quality X-ray observations of representative galaxy clusters. 
In this way we are able to generate realistic clusters and produce realistic X-ray full sky light-cones of dark matter only $N$-body simulations for \emph{eROSITA}. This empirical approach is complementary to phenomenological and semi-analytic models, as well as hydrodynamical simulations in that the modeled ICM profiles are non-parametric other than halo mass, redshift, and dynamical state. 

The structure of the paper is as follows: 
We discuss the cosmological N-body simulations used to create the mock X-ray light-cone catalogues and photons in Section~\ref{sec:nbody:data}. We describe models to simulate the X-ray properties of galaxy groups and clusters in Section~\ref{sec:cluster:model:xray}. 
Finally, in Section~\ref{sec:results} we discuss the results and predictions of the model. 

We assume a flat $\Lambda$CDM cosmology close to that of the \citet{Planck2014}, detailed in the following Section. 
The set of \emph{eROSITA} mocks is made public. Access is described in Appendix. 

%%%%%%%%%%%%%%%%%%%%%%%%%%%%%%%%%%%%
%
%
% section N-body
%
%
%%%%%%%%%%%%%%%%%%%%%%%%%%%%%%%%%%%%
\section{N-body data}
\label{sec:nbody:data}

\begin{table}
	\centering
	\caption{Characteristics and setup for each N-body dark matter only simulation. 
	L: length of the box [Gpc/h].
	$M_p$: Mass of the dark matter particle [${\rm M_\odot/}h$]. 
	$M^{cut}_{vir}$: Cut in halo mass [${\rm M_\odot/}h$]. 
	All haloes with $M_{vir}>M^{cut}_{vir}$ are included in the light-cone. The mass limit is rounded a little above 100 particles. 
	z$_{max}$: maximum redshift to the observer. 
	The minimum redshift is 0 for all light-cones. 
	N snap: Number of snapshots available within z$_{max}$.
	Ref: reference articles for the simulations (1): \citet{Klypin2016}, (2): \citet{Chuang2019}
	}
	\label{tab:light:cones}
	\begin{tabular}{lr cc rrr rr} 
	\hline
name             &  ${\rm L_{box}}$      & M$_p$    & ${\rm M^{cut}_{vir}}$     & z${\rm _{max}}$ &  N & Ref\\%& l o s stretch \\%& N haloes \\
                 & [Gpc/h] & [${\rm M_\odot/}h$] &   [${\rm M_\odot/}h$] & & snap & \\
\hline
\textsc{SMDPL }   & 0.4  & $9.63\times10^{7}$ & $1\times10^{10}$ & 0.43 & 41  & (1) \\ %      
\textsc{MDPL2 }   & 1.   & $1.51\times10^{9}$ & $2\times10^{11}$ & 6.2    & 91  & (1) \\ %
\textsc{HMDPL }   & 4    & $7.9\times10^{10}$ & $1\times10^{13}$ & 1.8  & 50 & (1) \\ %
\textsc{UNIT\_1  } & 1.   & $1.2\times10^{9}$  & $2\times10^{11}$ & 6    & 89  & (2) \\ %
\textsc{UNIT\_1i } & 1.   & $1.2\times10^{9}$  & $2\times10^{11}$ & 6    & 89  & (2) \\ %

\hline
	\end{tabular}
\end{table}

\begin{table*}
	\centering
	\caption{Statistics of (sub) haloes in each light-cone. 
	For five mass thresholds: $M_{vir}>M_{vir}^{cut}, 5, 7, 10, 50 \times 10^{13}$ $h^{-1} M_\odot$.  we count the number of halos and sub-haloes. 
	`h': number of haloes contained in the projected simulation.  
	`s': number of sub-haloes.
	}
	\label{tab:N:haloes:sat}
	\begin{tabular}{lrr rr rr rr rr rr} 
	\hline
name             & 	\multicolumn{10}{c}{N haloes (h: haloes, s: sub-haloes) with $M_{vir}>$} \\
                 & \multicolumn{2}{c}{ ${\rm M^{cut}_{vir}}$}  
                 & \multicolumn{2}{c}{ ${\rm 5 \times 10^{13}}$}  
                 & \multicolumn{2}{c}{ ${\rm 7 \times 10^{13}}$}  
                 & \multicolumn{2}{c}{ ${\rm 10^{14}}$} 
                 & \multicolumn{2}{c}{ ${\rm 5 \times 10^{14}}$} \\
                & h & s (s/d) & h & s (s/d) & h & s (s/d) & h & s (s/d) & h & s (s/d) \\
\hline
\textsc{SMDPL}  & 1,574,501,466 & 296,215,425 (18.8) &   319,746 & 10,796 (3.4)  &   192,165 &  5,840 (3.0) & 106,581 & 2,939 (2.8)  & 3,580 & 56 (1.6)  \\
\textsc{MDPL2}  & 5,711,284,943 & 534,492,535 (9.4)  & 2,219,880 & 40,903 (1.8)  & 1,124,378 & 18,200 (1.6) & 519,691 & 6,953 (1.3)  & 5,897 & 28 (0.5)  \\
\textsc{HMDPL}  &  28,897,265 &   1,150,831 (4.0)  & 2,046,611 & 35,339 (1.7)  & 1,051,562 & 14,491 (1.4) & 491,468 & 5,335 (1.1)  & 5,615 & 11 (0.2)  \\
\textsc{UNIT1}  & 5,923,863,122 & 545,028,953 (9.2)  & 2,115,664 & 38,174 (1.8)  & 1,060,239 & 16,573 (1.6) & 478,869 & 6,227 (1.3)  & 5,030 & 17 (0.3)  \\
\textsc{UNIT1i} & 5,917,731,143 & 544,387,870 (9.2)  & 2,120,236 & 37,946 (1.8)  & 1,069,035 & 16,470 (1.5) & 488,147 & 6,203 (1.3)  & 4,715 & 45 (1.0)  \\
\hline
\end{tabular}
\end{table*}

We use the MultiDark \citep[\textsc{HMDPL, MDPL2, SMDPL}][]{Klypin2016}\footnote{\url{https://www.cosmosim.org}} 
and the UNIT \citep[\textsc{UNIT1, UNIT1i}][]{Chuang2019}\footnote{\url{https://unitsims.ft.uam.es}} dark matter only simulations to create light-cones spanning the full sky, see Table \ref{tab:light:cones}. 
The MultiDark (UNIT) simulations are computed in a Flat $\Lambda$CDM cosmology with $H_0=67.77$ (67.74) km s$^{-1}$ Mpc$^{-1}$, $\Omega_{m0}=0.307115$ (0.308900), $\Omega_{b0}=0.048206$, $\sigma_8$=0.8228 (0.8147). 
Alternative simulations which could be used for this project include those of \citealt{Angulo2012,Skillman2014,Heitmann2015,Ishiyama2015,Ishiyama2020arXiv200714720I} as they have sufficiently high resolution for the AGN population to be accurately modeled, which requires the full population of haloes with mass greater than $10^{12}M_\odot$. 

The setup of each light-cone and parameters of each simulation box used (5 boxes in total) are detailed in Table \ref{tab:light:cones}. 
All simulations are consistently post-processed with the \textsc{rockstar} software \citep{Behroozi2013} to obtain haloes and sub-haloes. 
The halo mass function is correct to a few percent down to halo masses of $M_{vir}\sim10^{10}M_\odot$ for \textsc{SMDPL}, $M_{vir}\sim10^{11}M_\odot$ for \textsc{MDPL2, UNIT1, UNIT1i} and $M_{vir}\sim10^{13}M_\odot$ for \textsc{HMDPL}, see details in \citet{2017ComparatHMF,Chuang2019}. 

We consider haloes and sub-haloes
%more massive than $10^{13} M_\odot$ for \textsc{HMDPL}, $10^{10} M_\odot$ for \textsc{SMDPL} and $2\times 10^{11} M_\odot$ for \textsc{MDPL2, UNIT1, UNIT1i} \textit{i.e.} 
with slightly more than 100 bound particles per (sub) halo, see  Table \ref{tab:light:cones}. 
Table \ref{tab:N:haloes:sat} gives the number of halos and sub-halos for the complete halo population and for four mass thresholds: $M_{vir}>5, 7, 10, 50 \times 10^{13}$ $h^{-1} M_\odot$. 
There are about 2 million (40 thousand) haloes (sub-haloes) with a halo mass $M_{vir}>5 \times 10^{13}$ $h^{-1} M_\odot$ over the full sky. A quarter of them have $M_{vir}>10^{14}$ $h^{-1} M_\odot$ and only 2\% of them have $M_{vir}> 5 \times 10^{14}$ $h^{-1} M_\odot$. 
The comparison of the fraction of sub-halos in Table \ref{tab:N:haloes:sat} shows that the resolution matters to find sub-halos in dense environments. 
For comparison, the \emph{eROSITA} cluster sample is expected to be complete for masses greater than $10^{14} (7\times 10^{13}) M_\odot$ for redshifts $z<1$ ($z<0.3$) \citep{arxiv12093114_Merloni, Clerc2018AA...617A..92C}. 

\subsection{Light-cones}

From the list of snapshot output (${\rm z_{snap}}$) in each simulation, we compute the comoving distance between the snapshots redshifts and redshift 0, $d_C^{{\rm snap}} = d_C({\rm z_{snap}})$, using the exact cosmological parameters of the simulation: 
\begin{equation}
\label{eq:comoving:distance}
{\rm d_C(z)} = \int_0^z \frac{\mathrm{c} \; du}{\mathrm{H_0} \sqrt{ (1+u)^3 \Omega_{m0} + \Omega_{DE0} } } {\rm \;[Mpc]}.
\end{equation}
We compute the middle point distance between each snapshot. These middle points are the boundaries between the shells; i.e. the distances at which we switch between using the outputs in a snapshot to another. 
The array of boundaries read:
\begin{equation}
\begin{split}
& {\rm Mid\; points = array\left( \frac{d_C^{snap}[1:]+d_C^{snap}[:-1]}{2} \right)} \\ 
& {\rm d_C^{max\; boundary} = array\left( Mid\; points, d_C^{snap}[-1] \right) } \\
& {\rm d_C^{min\; boundary} = array\left( d_C^{snap}[0], Mid\; points \right) }. 
\end{split}
\end{equation}
% & {\rm d_C^{mid}_{,i} = \left( \frac{d_C(z_i)+d_C(z_{i+1})}{2} \right)} \\ 
% & {\rm d_C^{max\; i} = array\left( Mid\; points, d_C^{snap}[-1] \right) } \\
% & {\rm d_C^{min\; i} = array\left( d_C^{snap}[0], Mid\; points \right) }. 

We compute the maximum number of replicas needed for each snapshot:
\begin{equation}
{\rm  N_{replicas} = int( d_C^{max\; boundary} / L_{box} ) + 1 }.
\end{equation}

We replicate all snapshots available ${\rm (2\; N_{replicas})^3}$that are within the redshift range of interest. 
The maximum to reach redshift 6 is 12$^3$=1728 replicas. 
We use the periodic boundary condition to have consistent large scale structure across boundaries. 
It also preserves the angular (2D) and monopole (3D) clustering information on the full sky \citep[e.g.][]{Lindholm2020arXiv201200090L}. 
Note that the accuracy of the correlation functions measured on the replicated mock is limited by the volume of the initial simulation. These mocks are not suited to estimate uncertainties on clustering.
We obtain a regular grid with the observer located at (0,0,0). In each replica, the coordinates (${\rm x_{\rm snap}, y_{\rm snap}, z_{\rm snap}}$) are shifted by units of box length: 
\begin{equation}
\begin{split}
{\rm  x} & \longleftarrow {\rm x_{\rm snap} + i \times L_{box} \;[Mpc]},  \\ 
{\rm  y} & \longleftarrow {\rm y_{\rm snap} + j \times L_{box} \;[Mpc]},  \\ 
{\rm  z} & \longleftarrow {\rm z_{\rm snap} + k \times L_{box} \;[Mpc]},  
\end{split}
\end{equation}
where (i,j,k) takes values between -${\rm  N_{replicas}}$ and ${\rm  N_{replicas}-1}$. 
In each replicated snapshot, we select objects with a comoving distance to the observer between ${\rm d_C^{min\; boundary}}$ and ${\rm d_C^{max\; boundary} }$.
We concatenate each set to obtain a full-sky shell.  
The width of the shells varies between 100 and 120 Mpc, except for SMDPL where it is between 20 and 60Mpc. 
The properties of objects in a shell are at a fixed time (that of the snapshot). The redshifts of each (sub) halo (computed with or without line-of-sight velocity projection) give an accurate position, but not an accurate proper time, defined by the snapshot redshift. 
Indeed, we do not interpolate the positions and the velocities between snapshots to obtain a more continuous redshift sampling and set of halo properties. 
It is not needed in this study. 
The replication of snapshot and extraction of the shell typically requires 20-60 CPU hours. 

As pointed out by \citet{Blaizot2005MNRAS.360..159B} and \citet{Merson2013MNRAS.429..556M}, when replicating the simulations artefacts appear due to the alignment of structures. 
As we simulate the full sky, we cannot apply the technique proposed by \citet{Blaizot2005MNRAS.360..159B}, best suited for pencil beam surveys. 
For \textsc{HMDPL}, the simulation is sufficiently large that there is no such artifact. 
For the 1Gpc/h simulations, the replications of a structure at redshift 0.1 (age 12.48 Gyr) do occur at redshifts 0.1 , 0.49, 0.99, 1.71, 2.82, 4.71
( or ages 8.71,  5.91,  3.82,  2.31,  1.27 Gyr). 
These redshifts are more spaced in time than the dynamical times of the haloes. 
We find that the effects shown in Fig. 1 and 2 of \citet{Blaizot2005MNRAS.360..159B} are hardly visible until redshift 2. 
Above redshift 2, as the number of replica increases drastically, they become visible. 
Thus the \emph{eROSITA} cluster sample located below $z<1.5$ is unaffected by these artefacts. 
The \emph{eROSITA} AGN sample with an extended high redshift tail is affected when considering the full sky area. 
For the smallest simulation, \textsc{SMDPL}, we limit the simulation in redshift to have a maximum of two line-of-sight replications. Due to is smaller volume and redshift reach, the use of \textsc{SMDPL} will be limited to studying the galaxy population (sub-haloes) in low redshift groups and clusters. 
The \textsc{SMDPL} simulation will not be used to create full sky X-ray maps where these artefacts matter. 

Using \textsc{SMDPL}, we create one low redshift ($z<0.43$) light-cone that resolves groups and clusters and the sub-haloes therein. 
Using \textsc{HMDPL}, we create a cluster only light-cone as its resolution is too coarse to add AGN or galaxies in clusters. 
Using the \textsc{MDPL2, UNIT1, UNIT1i} simulations, we create light-cones ($z<6$) that contain all clusters, their galaxies (with some degree of incompleteness at low redshift) and the large scale structure sampled by AGN. 
The construction of the light-cone is close to that used by \citet{Merson2013MNRAS.429..556M}. 
In the \textsc{MDPL2, UNIT1}, and \textsc{UNIT1i} light-cones, 
thanks to a sufficiently high resolution and a large volume, we simulate both the AGN and the clusters \citep{2018MNRAS.tmp.3272G,Comparat2019MNRAS.tmp.1335C}. 
Note that due to the limited volume, the high mass end ($>10^{15}M_\odot$) of the halo mass function remains noisy. 
Below redshift $z<0.1 - 0.2$ the galaxies in the clusters and AGN populations suffer from the limited resolution of the simulation.  

\subsection{Coordinates}
\label{sec:coordinates}
For each halo, we compute the angular coordinates (equatorial, galactic, ecliptic) and line-of-sight velocity vector to infer the redshift in real and redshift space. 
The simulated positions ($x$, $y$, $z$) and peculiar velocities ($v_x$, $v_y$, $v_z$) enable computation of observed coordinates as follows
\begin{equation}
d_C=\sqrt{x^2 + y^2 + z^2} {\rm \;[Mpc]}.
 \end{equation}
From $d_C$, we infer the real space redshift: \texttt{`redshift\_R'} through the comoving distance -- redshift relation given in Eq. \ref{eq:comoving:distance}.
Angular coordinates (in degrees) are then computed:  
\begin{equation}
\begin{split}
&{\rm \theta  = arccos( z / d_C )  180/\pi\; [^\circ],} \\
&{\rm \phi = arctan2( y, x )  180/\pi\; [^\circ],} \\
&{\rm R.A. = \phi + 180\; [^\circ],} \\
&{\rm Dec. = \theta - 90\; [^\circ],} 
\end{split}
\end{equation}
and converted to galactic and ecliptic coordinates with \textsc{astropy}. 
For convenience, we split the sky into 768 equal are pixels using \textsc{healpix} (NSIDE$=8=2^3$ in the nested scheme) where the colatitude is $90-{\rm Dec.}$ and the longitude is R.A. \citep{Gorski2005ApJ...622..759G}. 
We project the peculiar velocity along the line-of-sight as follows:
 \begin{equation}
v_\| = (v_x . x + v_y . y + v_z . z ) / {\rm d_C \; [km \; s^{-1}]},
\end{equation}
and obtain the comoving distance with peculiar motion projected as:
\begin{equation}
{\rm d_C^s = d_C} + v_\| / H(z_R){\rm \; [Mpc]}.
\end{equation}
From ${\rm d_C^s}$ and Eq. \ref{eq:comoving:distance}, we infer the observed redshift: \texttt{`redshift\_S'}. 

\subsection{Foreground absorption}
Given each (sub) halo's position on the sky, we retrieve the Milky Way HI column density from \citet{HI4PI2016AA594A116H} and the E(B-V) extinction value taken from \citet{2014AA...571A..11P} using the \textsc{dustmap} software \citep{2018JOSS....3..695M}. These values are later used to attenuate fluxes.

%%%%%%%%%%%%%%%%
%
%
% X ray model
%
%
%%%%%%%%%%%%%%%%
\section{X-ray cluster empirical model}
\label{sec:cluster:model:xray}

\begin{figure}
\centering
\includegraphics[width=.9\columnwidth]{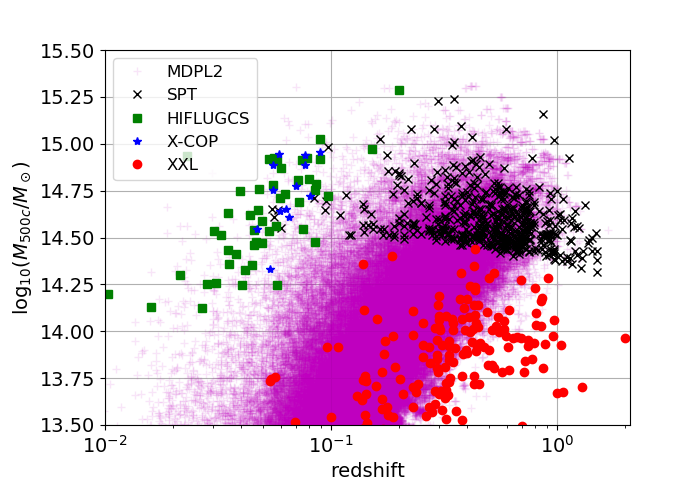}
\caption{\label{fig:mass:redshift}$\log_{10}(M_{500c}/M_\odot)$ vs redshift. 100,000 clusters with the highest flux in the full sky light-cone extracted from \textsc{MDPL2} simulation (lila transparent crosses). 
Including fainter clusters would fill the lower-right part of the diagram. 
The sample used to create the covariance matrix are shown with the symbols specified in the legend (XXL: red circle, SPT: black cross, HIFLUGCS: green square, X-COP: blue star). }
\end{figure}

The goal here is to create a model that represents the overall population of clusters and groups to be detected by \emph{eROSITA}. 
A description of its selection function is available in \citet{Clerc2018AA...617A..92C}. 
The \emph{eROSITA} cluster sample is expected to be complete for masses greater than $10^{14} (7\times 10^{13}) M_\odot$ for redshifts $z<1$ ($z<0.3$).
Thus, we consider the (sub) halos with a mass $M_{500c} > 5\times 10^{13}$M$_\odot$ ($\log_{10}(5\times 10^{13})=13.7$). 

There exists a number of approaches to paint galaxy clusters in the X-ray, 
for example,  
phenomenological modelling of cluster pressure and temperature profiles with dependence on dynamical state \citep[\textit{e.g.}][]{Zandanel2018MNRAS.480..987Z},  
or semi-analytical, physically informed relations \citep[\textit{e.g.}][]{Shaw2010ApJ...725.1452S,Flender2017ApJ...837..124F}, 
or machine learning algorithms trained on hydro-dynamical simulations \citep[\textit{e.g.}][]{Cui2018MNRAS.480.2898C}. 
The model created here could be regarded as an upgrade of the first method.

\subsection{Surface brightness profiles}
\label{subsec:surf:brigh:prof}

\begin{figure}
\centering
\includegraphics[width=.9\columnwidth]{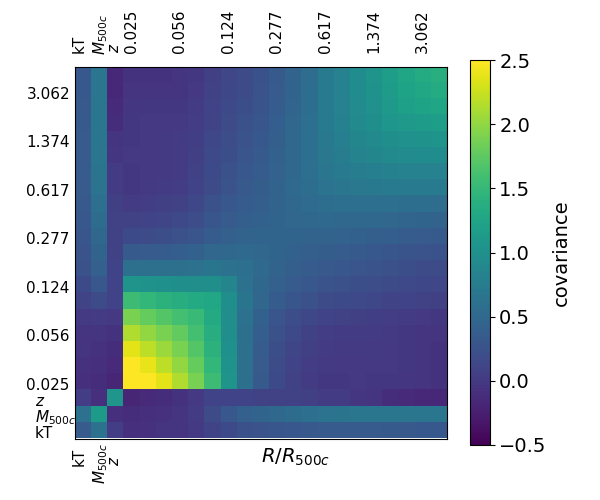}
\includegraphics[width=.9\columnwidth]{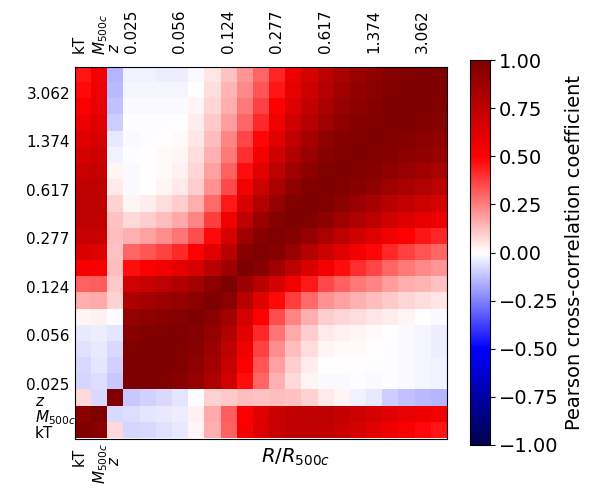}
\caption{\label{fig:cluster:covariance:matrix}
Covariance matrix (top) and Pearson's cross-correlation coefficient (bottom) derived from the data. 
}
\end{figure}

\begin{figure}
\centering
\includegraphics[width=.9\columnwidth]{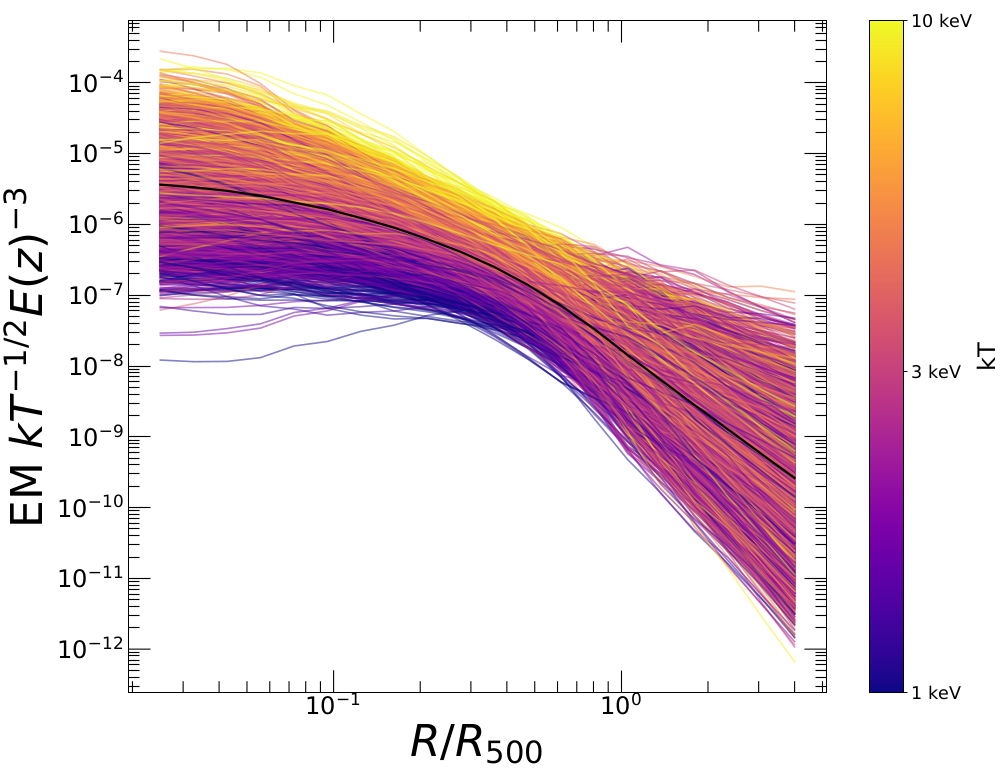}
\caption{\label{fig:cluster:profiles:used}
Model emissivity profiles [unit: ${\rm cm^{-6}\; Mpc \; {\rm keV}^{-1/2}}$] defined in Eq. \ref{eqn:EMr} scaled by temperature and $E(z)$ (see Eq. \ref{eq:profile}) as a function of radius scaled by $R_{500c}$ (see Eq. \ref{eq:r500c}). 
The correlation with temperature seen in the covariance matrix is present in the simulated profiles. 
}
\end{figure}

\begin{figure}
\centering
\includegraphics[width=.9\columnwidth]{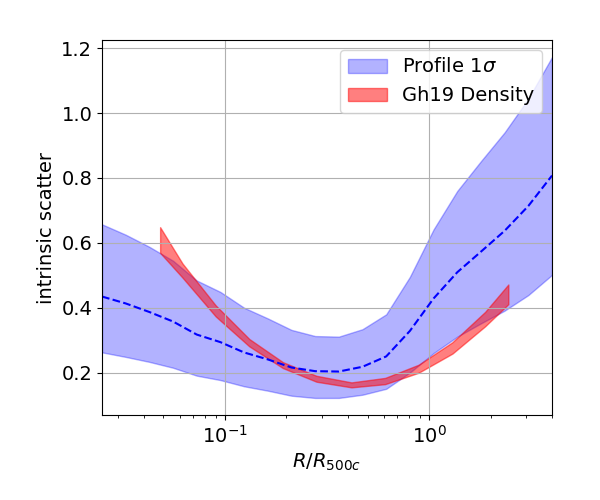}

\caption{\label{fig:cluster:profiles:scatter}
Intrinsic scatter as a function of scale. The blue shaded area shows the 32nd and 68th percentiles of the distribution of the log10 of the profiles divided by the mean profile. 
We show the intrinsic scatter measurements of the density profiles from \citet[Gh19, red]{Ghirardini2019}. 
The agreement with the shape found by Gh19 is good.
The Gh19 data is of better quality (deeper) than the data used to calibrate the method. 
This shows that our method slightly underestimates the scatter on small scales.  
}
\end{figure}

To simulate realistic surface brightness profiles, 
we start from a collection of measured profiles from several existing samples spanning a wide range in mass and redshift, as shown in Fig. \ref{fig:mass:redshift}. 
The considered samples are XMM-XXL \citep{Pierre2016,Adami2018A&A...620A...5A}, 
HIFLUGCS \citep{Reiprich2002}, 
X-COP \citep{Eckert2019} 
and SPT-Chandra \citep{Sanders2018MNRAS}. 
XXL is a serendipitous X-ray survey over an area of 50 square degrees. 
The detected clusters span a redshift range of $z=(0.05-1.1)$ and mainly populate the galaxy group and low-mass cluster regime, with a median mass  of $\sim10^{14} M_\odot$. 
As such, it is quite similar to the \emph{eROSITA} survey. 
HIFLUGCS is a complete sample of the brightest clusters detected in the \emph{ROSAT} all-sky survey and constitutes a good low-redshift anchor. 
X-COP was selected from the strongest Sunyaev-Zeldovich detections in the \emph{Planck} survey and thus samples the massive end of the cluster population. 
The SPT sample was drawn from a sample of the 90 most massive SPT-SZ detections at $z>0.3$, for which a homogeneous \emph{Chandra} follow-up program was conducted \citep{McDonald2013,McDonald2014}. 
Altogether, our sample comprises a set of 322 (182 + 45 + 12  + 83, XXL + HIFLUGCS + X-COP + SPT) profiles. 
For the details of the data analysis and profile extraction methods, we refer to \citet[][XMM-XXL]{Eckert2016}, \citet[][HIFLUGCS sub-sample limited to $kT\geq 3$ keV clusters]{Kaefer2019}, \citet[][X-COP]{Ghirardini2019} and \citet[][SPT]{Sanders2018MNRAS}. 
The mass and redshift ranges covered by these surveys encompass most of the parameter space that will be probed by \emph{eROSITA}. 
The dataset will eventually be complemented with measurements from the \emph{eROSITA} survey itself.
 
For each cluster, we additionally have measurements of their redshift, spectroscopic temperature, and mass. 
The XXL masses are based on a mass-temperature relation calibrated through Subaru/HSC weak lensing measurements of 136 XXL clusters \citep{Umetsu2020}. 
The X-COP masses are drawn from high-quality hydrostatic mass measurements from combined \emph{XMM-Newton} and \emph{Planck} data and are found to be consistent with weak-lensing measurements for a subset of systems \citep{Ettori2019,Eckert2019}. 
The HIFLUGCS masses are derived from the \emph{Planck} $Y_{SZ}-M_{500c}$ relation \citep{PSZ2} and are corrected for a fiducial hydrostatic bias of 20\% for consistency with the other datasets.
The SPT masses are determined using the best-fit scaling relation (4 parameters) for a flat $\Lambda$CDM cosmology with $\Omega_m$=0.3, h=0.7 and $\sigma_8$=0.8, see \citet{Bocquet2019ApJ...878...55B} for more details. 
Fig. \ref{fig:mass:redshift} shows how the calibration data set compares to the simulated lightcone in the mass redshift plane. We find that the set of clusters used for calibration covers the relevant parameter space, except for the low mass and low redshift clusters and groups, see discussion at the end of this Section.

\subsubsection{Emissivity profiles}

To create a consistent library of profiles from the initial heterogeneous data set, we scale each profile according to the self-similar mass and redshift evolution \citep{Neumann1999AA,Arnaud2002} and interpolate them onto a common grid of 20 logarithmic-spaced points in the radial range of $R=[0.02-4]R_{500c}$. 

When the outskirts of the detected clusters are not covered by the X-ray data, we extrapolate their profile with a power-law. 
The slope of the profile is estimated from the outermost 3 data points, and the emissivity beyond the maximum detection radius is assumed to follow a power law smoothly connected to the outermost data point. 
In this analysis, luminosity and flux are estimated by integrating the profiles up to $R_{500c}$, while events are simulated up to 2$R_{500c}$. 
In observations,the cluster surface brightness is dominated by that of the background  at $R>R_{500c}$. 
So, even if the profile extrapolation method adopted is not the most accurate, it will hardly be visible (blended with noise) in the event maps for this simulation setup. 
Thus, these mock clusters may not be useful for studying cluster outskirts. 

Within $R_{500c}$, we investigate possible biases. 
We create four mean vector and covariance matrices (see next section) with extrapolation slopes biased by +20, -20, +40, and -40 \%. 
We create full sky mock catalogues, so the comparison is not limited by statistics. 
For $M_{500c} > 7\times10^{13}M_\odot$, we measure a relative change in the mean of the mass--luminosity ($M_{500c}$-- $L_X^{500c}$) scaling relation smaller than $\pm0.5\%$. The effect of the extrapolation are negligible and thus we find the method robust enough for our purpose. 

The emission measurement (or emissivity) along the line-of-sight as a function of radius, $EM(r) = \int n_e n_H dl$, is defined as in \citet{Neumann1999AA} and \citet{Arnaud2002}:  
\begin{equation}
\label{eqn:EMr}
{\rm EM(r)}\;  {\rm [cm^{-6}\; Mpc]} = \frac{4\pi(1+z)^4 S(r / d_A(z))}{\epsilon(T, z)}. \end{equation}
The unit ${\rm [cm^{-6}\; Mpc]}$ is used for convenience. 
EM$(r)$ is proportional to the ratio of the X-ray surface brightness profile, $S(\theta)$ ${\rm [ct\; s^{-1}\; arcmin^{-2}]}$, and the emissivity of the instrument, $\epsilon(T,z)$ [ct s$^{-1}$ cm$^{5}$].
$\epsilon$ is the integral over the energy band (depending on the survey considered) of the detector effective area [cm$^{2}$], times an inter-stellar absorption term, times the emissivity of a plasma of temperature $T$, heavy element abundance A, and redshift z [ct s$^{-1}$ cm$^{3}$ keV$^{-1}$]. 
The model emissivity for each cluster is assumed isothermal. 
The angular scale ($\theta$) and physical scale ($r$) are related via the angular diameter distance: $r = \theta d_A(z)$. 

\subsubsection{Covariance matrix}
We construct the mean vector of the data: redshift, temperature, masses M$_{500c}$ and the emissivity as a function radius (20 points, normalized by $R_{500c}$) (23 elements). 
We then construct a covariance matrix between these quantities, see Fig. \ref{fig:cluster:covariance:matrix} 
(the resulting matrix has a shape 23 x 23, the Pearson's cross-correlation coefficient is also shown). 
The positive correlation between mass and emissivity as a function of radius is related to the deviation from the self similar model caused by the gas fraction. 
For the same reason, but to a lesser extent, $kT$ correlates with emissivity as a function of radius. 
The anti-correlation between redshift and mass is a selection effect, see Fig \ref{fig:mass:redshift}. 
The covariance between the emissivity values as a function of radius are expected at small radius, see Fig \ref{fig:cluster:profiles:scatter}. 
At large radius, the covariance might be over-estimated. 
The correlation between $kT$ and mass is illustrated in Fig. \ref{fig:cluster:scaling:relation:kT:M500c}, that shows the scaling relation.

Through a Gaussian multivariate random process (using the log of the mean vector and the covariance matrix i.e. values are log-normally distributed), we generate a set of simulated clusters characterized by their X-ray emissivity profile, redshift, mass and temperature. 
Fig. \ref{fig:cluster:profiles:used} shows an example set of simulated cluster profiles. 
The simulated profiles are color-coded by the temperature. 
We see that our procedure captures the known mass dependence of the X-ray emissivity profiles. Group-scale objects show a lower central emissivity and a flatter profile than cluster-scale systems \citep{Eckert2016}. 
We predict an intrinsic scatter of the emissivity as a function of scale that is in agreement with that of the density profiles measured by \citet{Ghirardini2019}, see Fig. \ref{fig:cluster:profiles:scatter}.

\subsubsection{Derived quantities}

From the redshift and the halo mass, we deduce for each cluster the radius corresponding to 500 times the critical density, $\rho_c$, as follows:
\begin{equation}
\label{eq:r500c}
r_{500c} = \left(\frac{3 M_{500c}}{ 4 \pi 500 \rho_c(z)  }\right)^{1/3} {\rm[ cm ]}.
\end{equation}
The iso-thermal surface brightness profile, $\mathrm{profile}(x)$, is obtained by :
\begin{equation}
\label{eq:profile}
\mathrm{profile}(x) = {\rm EM(x)}  \sqrt{kT/[10 {\rm keV}]} E^3(z) \Lambda(T), {\rm[erg\; s^{-1} cm^{-2}]},
\end{equation}
where $x=r/r_{500c}$ and $\Lambda(T)$ is the cooling function [cm$^3$ erg s$^{-1}$] in the band 0.5-2~keV \citep{Sutherland1993ApJS...88..253S}. 
%It takes values of 8.5e-24 (5.5e-24), at 1 (2) keV. 
The luminosity at $x$ is obtained with the cumulative sum (equivalent to integrating over a cylinder) :
\begin{equation}
\label{eq:lx:integral}
L_X(r) = \int^r_{x=0} \left( \mathrm{profile}(x) r^2_{500c} 2 \pi x dx \right)\; {\rm[erg\;  s^{-1}]}.
\end{equation}
For each simulated cluster, we record its redshift, temperature, mass, X-ray luminosity in the band 0.5-2~keV and the surface brightness profile in an image format compatible with the \textsc{sixte} simulator \citep{Dauser2019AA...630A..66D}, see Sec. \ref{subsec:images}.

\begin{figure}
\centering
\includegraphics[width=.9\columnwidth]{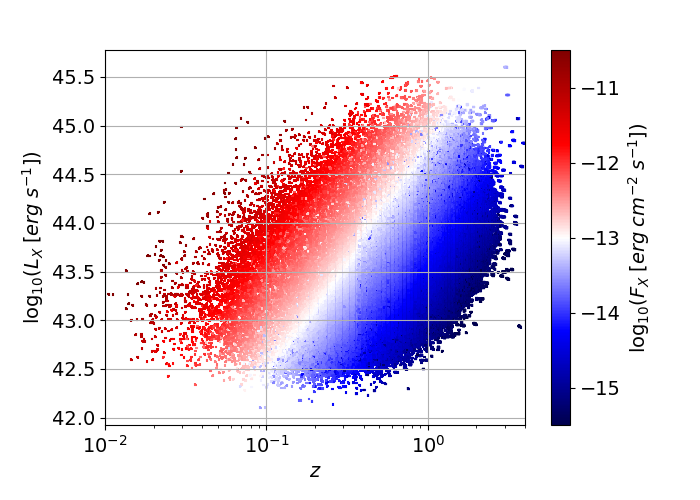}
\includegraphics[width=.9\columnwidth]{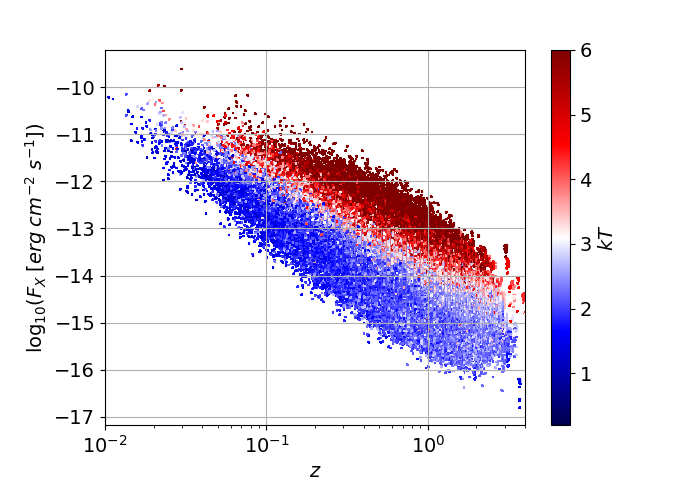}
\caption{\label{fig:k:correction} \textbf{Top.} Luminosity (rest-frame soft X-ray band) as a function of redshift colored according to flux (observed frame soft X-ray band). 
\textbf{Bottom.} Flux as a function of redshift colored according to temperature.  
}
\end{figure}

\subsection{Linking simulated profiles to dark matter haloes}

In the dark matter only N-body simulation, each halo (sub-halo) is characterized by its redshift and mass (M$^{DM}_{500c}$). 
The dynamical state of a halo is traced by {\tt $X_{\rm off}$}, defined as the distance between the halo highest density peak and its centre of mass divided by its virial radius. 
\citet{Klypin2016} classified halos as relaxed or disturbed using a combination of Spin and {\tt $X_{\rm off}$}. 
Typically, relaxed haloes have {\tt $X_{\rm off}\in[0.01,0.07]$} and disturbed haloes have  {\tt $X_{\rm off}\in[0.07,0.3]$}. 
\citet{Henson2017MNRAS.465.3361H} discussed possible criteria to describe the relaxation state of the cluster: \textit{e.g.} substructure fraction, spin parameter, etc and retain {\tt $X_{\rm off}$} as the cleanest (and most direct) estimator. 

To ensure the availability of simulated profiles from each light-cone shell, we generate 100 times more profiles from the covariance matrix than the number of clusters present in a given shell. 
In this way, we densely sample the parameter space, avoiding using the same profile twice. 
We assign to each halo (and sub-halo) a simulated cluster using a nearest neighbour procedure in three dimensions. 
The near-neighbour search for redshift is direct.  
Total mass (including baryons) and dark matter mass (in a collisionless simulation) differ as a function of mass due to baryonic effects \citep[e.g.][]{Velliscig2014MNRAS.442.2641V,Cui2016MNRAS.458.4052C}.
These mock catalogues allow to shift the X-ray appearance of halos from its nominal relation to the total mass. 
This can be used to study the effects of mass calibration and the effects of baryonic deficiency on recovering the cosmological parameters.
We introduce a mass bias parameter, $b_{M}$, and look for the nearest neighbour between M$^{\rm clusters}_{500c}$ and $b_{M}$M$^{DM}_{500c}$. 
Because we focus on haloes with M$^{DM}_{500c}>10^{13.7}$, a constant $b_{M}$ should absorb such effects. If the method were to be generalized to lower masses (groups), $b_{M}$ should become mass dependent. 
This is the only parameter of this model, set to 1 (unless stated otherwise). 
This parameter induces significant differences when predicting the density of clusters as a function of their brightness i.e. in the luminosity function and in the logN-logS. 
In Section \ref{subsection:number:density}, we show simulation results for $b_{M} \in [0.6,1.1]$, a range similar to that used by \citet{Zandanel2018MNRAS.480..987Z}. 

Since the shape of the surface brightness profile is related to the relaxation state of the halo, we preferentially assign flatter profiles to the most disturbed systems, according to their {\tt $X_{\rm off}$} value. 
We relate the negative $\log_{10}$ of the central emissivity to {\tt $X_{\rm off}$}. 
\begin{equation}\label{eq:xoff:em0:link}
 - \log_{10}({\tt EM(r=0)}) \propto X_{\rm off}. 
\end{equation}
A low {\tt $X_{\rm off}$} (a high ${\rm EM(0)}$) corresponds to a cool-core (relaxed) cluster. 
A high {\tt $X_{\rm off}$} (a low ${\tt EM(0)}$) corresponds to a non cool-core (un-relaxed) cluster. 
{\tt $X_{\rm off}$} and ${\tt EM(0)}$ have quite different face values, so for the nearest-neighbour match to work, we normalize them. 
We compute the cumulative distribution function for {\tt $X_{\rm off}$} and ${\tt EM(0)}$. The cumulative distribution function relates a value (of {\tt $X_{\rm off}$} and ${\tt EM(0)}$) to its percentile in its parent distribution function. 
The nearest neighbour is then chosen with the percentiles of both quantities.  
Note that relating ${\tt EM(0)}$ to {\tt $X_{\rm off}$} does not skew or change the intrinsic population of simulated clusters i.e. {\tt $X_{\rm off}$} is not used to preferentially select certain types of clusters over others. 
Instead, among the simulated clusters (a fair sample), it relates ${\tt EM(0)}$ to {\tt $X_{\rm off}$} after ordering them. 
In this manner, the cool-core bias is physically related to the halo properties and possibly its surrounding large-scale structure \citep{Eckert2011AA...526A..79E}. 
We force a relation to a parameter of the dark matter halo to possibly help reconstructing the selection function in a continuous fashion \citep[e.g.][]{Kaefer2019,Kaefer2020,Seppi2020arXiv200803179S}. 
Note that the true distribution of {\tt EM(0)} is not well known. 
Nevertheless, \citet[][Fig. 2]{Rossetti2017MNRAS.468.1917R} shows a distribution of the concentration of Planck clusters, which is similar to a log-normal distribution. \citet{Nurgaliev_2017} reached similar conclusions based on the SPT sample. 
In the future, with more constraints from observations, we will be able to better model this distribution.

As the model sample is finite, the precision of the nearest neighbour match will depend on its total size. 
The fractional differences in masses are smaller than 2\% for $M^{DM}_{500c}>5\times10^{13}M_\odot$. 
The fractional differences in redshift are smaller than 2\%. 
At low redshift and/or at low masses, we see the limitations of the sample we draw from. 
Indeed, the original sample from which simulated clusters are drawn does not contain enough clusters to smoothly cover the full mass-redshift parameter space of interest. 
Modeling more clusters at low masses and low redshift does not shrink the differences significantly. 
The low mass and low redshift regime will benefit from \emph{eROSITA} observations and should improve over the coming years, see Fig \ref{fig:mass:redshift}. 
The model would benefit from using more observations of low mass, low redshift clusters and groups e.g. the eeHIFLUGCS sample \citep[][]{Reiprich2017xru..conf..189R}. %(eeHIFLUGCS for example)

At the end of this procedure, each DM halo in the light-cone is linked to a surface brightness profile, a temperature and an X-ray luminosity.

\subsubsection{Limitations}

We verify that the mean of ${\tt EM(0)}$ in bins of haloes mass does not depend on haloemass. 
A linear fit gives a correlation coefficient of $-0.013$, indicating no significant dependence on the halo mass.  
We find that the mean of ${\tt EM(0)}$ in bins of temperature depends slightly on the temperature. 
A linear fit gives ${\tt EM(0)} = -0.7 \log_{10}(kT) + 5.2$ with a correlation coefficient of 0.17. 

Although at times symmetric (low  $X_{\rm off}$) clusters have a high central surface-brightness (high ${\tt EM(0)}$), it is not always the case that a high central surface brightness indicates symmetry for a cluster. 
For example \citet[][Fig. 2]{Mantz2016MNRAS.456.4020M} showed that at a given over-density, there is intrinsic scatter in the surface-brightness. 
In this approach, the scatter is present by construction, but its correlation (via Eq. \ref{eq:xoff:em0:link}) to the $X_{\rm off}$ parameter is tighter than it should be. 
With upcoming eROSITA observations, we will hopefully constraint the relation and its scatter (currently assumed to be zero) between relaxation state and central emissivity, to improve on Eq. \ref{eq:xoff:em0:link} \citep{Seppi2020arXiv200803179S}.

\subsection{Flux}

We deduce observed X-ray luminosities and fluxes in the soft X-ray band 0.5-2~keV (like \emph{eROSITA}) using the \textsc{xspec} software combined with an APEC\footnote{\url{http://www.atomdb.org}} spectrum extincted by a TBABS model. The metallicity is assumed to be 0.3 solar \citep{Asplund2009ARA&A..47..481A}. 
K-correction and attenuation are pre-computed on a large grid of redshifts, temperatures and hydrogen column density (n$_H$), then interpolated and applied to the mock clusters. 
Fig. \ref{fig:k:correction} shows the conversion from luminosity to flux and its dependence on redshift and temperature. 
Finally, fluxes are extincted using the local n$_H$ value and we obtain an `observed' mock catalogue. 

\subsection{Images}
\label{subsec:images}
\citet{Oguri2010MNRAS.405.2215O} measured, with the weak lensing technique, the ellipticity of dark matter haloes hosting galaxy clusters. 
This measurement is in excellent agreement with the standard collisionless cold dark matter model. 
\citet{Umetsu2018ApJ...860..104U} found that the ellipticity of the X-ray emission by the ICM in clusters is correlated to the dark matter halo ellipticity, suggesting a tight alignment between the intracluster gas and dark matter. 
So, having elliptical 2D surface brightness is key for upcoming studies characterizing their detection and selection function. 
For each model cluster, using the ellipticity of the dark matter halo and the model profile, we create an image (\textsc{simput} format, \citealt{Dauser2019AA...630A..66D}). 

We use the axis ratio fitted on the 3D dark matter halo spheroid ({\tt b\_to\_a\_500c}) as the axis ratio on the sky. 
Because {\tt $X_{\rm off}$} and the axis ratio are correlated \citep[e.g.][]{Lau2020arXiv200609420L}, the profile shape ${\tt EM(0)}$ becomes correlated to the ellipticity of the X-ray surface brightness contour. 
Note that the ellipticity in the simulation is determined on the mass density and not on the potential, which is traced by the X-ray emitting gas \citep[e.g.][]{Lau2011ApJ...734...93L}. 
Nevertheless, the Quartiles (Q1, 2, 3) of the ({\tt b\_to\_a\_500c}) axis ratio distribution: 0.45, 0.55 (median), 0.65, are in good agreement with observations \citep[e.g.][]{Shin2018MNRAS.475.2421S}. 
So even if the physical link is not direct, the face values are following a reasonable distribution. 
We find that using an axis ratio estimated on a larger aperture (e.g. 200c) leads to a set of clusters that would be too elliptical. 

%We defer to future work to extend this method to create weak lensing observables \citep{Giocoli2016MNRAS.461..209G, Umetsu2020arXiv200700506U}. 

\subsection{Sub-halos}
We also model sub-halos to reproduce substructure in the largest halos. 
Massive sub-haloes are rare, only a few percent of most massive halos contain a very massive substructure. 
Table \ref{tab:N:haloes:sat} gives the number of halos and sub-halos for four mass thresholds: $M_{vir}>5\times 10^{13}$, $7 \times 10^{13}$,  $5 \times 10^{14}$, $5 \times 10^{14}M_\odot/h$. 
The fraction of sub-halo depends on mass resolution i.e. on the ability to resolve a substructure within a very dense environment. 
The fraction of sub-halo with $M_{vir}>7 \times 10^{13}M_\odot/h$ varies from 1.4 (HMDPL, lowest resolution) to 3 (SMDPL, highest resolution) per cent. 
The modelling of sub-haloes becomes important at lower masses when emission becomes clumpy \citep{Osmond2004MNRAS.350.1511O}. It is left for future studies to push this method to the group scale.

\subsection{Other models}

We compare our empirical model with a phenomenological and a semi-analytic model in the literature. 

\subsubsection{Phenomenological model by \citet{Zandanel2018MNRAS.480..987Z}}

\citet[hereafter Za18]{Zandanel2018MNRAS.480..987Z} developed a phenomenological model, starting from the Planck pressure profiles \citep{Planck2013AA...550A.131P,Planck2016AA...594A..24P} and Chandra temperature profile models adapted from \citet{2010A&A...513A..37H} and including four different dynamical states for clusters, to implement ICM properties onto dark-matter-only halos. 
They use the MultiDark \citep{Klypin2016} simulation suite to create a set of \emph{eROSITA} mock light cones, available here\footnote{\url{http://skiesanduniverses.org/}} \citep{skies_universes_2017arXiv171101453K}. 
This phenomenological model reproduces well state-of-the-art
X-ray and SZ observations of galaxy clusters.

\subsubsection{Semi-analytic model by \citet{Shaw2010ApJ...725.1452S}}\label{sec:Sh10}

The semi-analytical model of the ICM presented in \citet[hereafter Sh10]{Shaw2010ApJ...725.1452S}, which was based on \citep{Ostriker2005ApJ...634..964O}. It assumes that for each halo, DM density profile follows the NFW profile which is uniquely determined by halo mass and halo concentration. For this paper we use the mass-concentration relation from \citet{Diemer2014ApJ...789....1D}. The gas pressure is in hydrostatic equilibrium with the NFW gravitational potential. The gas density and the gas pressure is related by a polytropic equation of state with polytropic index of $\Gamma = 1.2$ outside the cluster core $\geq 0.2 R_{500c}$ as suggested by observations \citep[e.g.][]{Ghirardini2019AA...627A..19G}. The gas density and total pressure profiles are determined by solving the equations of energy and momentum conservation.  Star formation, feedback from supernovae and active galactic nuclei are included in the model. In \citet{Shaw2010ApJ...725.1452S}, the model is extended with the inclusion of non-thermal pressure.  The model is further extended with the inclusion of cool-core, where we model a different polytropic index in the cluster core \citet{Flender2017ApJ...837..124F}, and gas density clumping in \citet{Shirasaki2020MNRAS.491..235S}. The model parameters used in the current paper was calibrated with density profiles measured with \emph{Chandra} observations of galaxy clusters detected by the South Pole Telescope (SPT) and with $M_{\rm gas}-M$ relations from \emph{Chandra} and \emph{XMM-Newton} observations. We refer the reader to \citet{Flender2017ApJ...837..124F} for details on the calibration. Note that clumping is not included in the current calibration. Also note that the current model does not include intrinsic scatter in the ICM profiles, meaning any two halos with the same mass and redshift will have the same ICM profiles. Intrinsic scatter in the ICM profiles due to the halo formation histories and halo shapes will be implemented in the future. 

\section{Results}
\label{sec:results}

We obtain scaling relations and their scatter from the generated mock catalogues, these are discussed in Section \ref{subsec:scaling:relations}. 
In Section \ref{subsection:number:density} we present the number of model clusters per square degrees as a function of flux. 
Finally, in Section \ref{subsec:fullsky:photons} we discuss the distribution of model photons as they will be observed by \emph{eROSITA}.

\subsection{Scaling relations}
\label{subsec:scaling:relations}

We measure a set of scaling relations between the halo properties and X-ray properties from the generated model catalogues. 
We consider an aperture corresponding to the 500c over-density and a soft band in the X-ray: 0.5-2~keV. 
The scaling relations obtained are close (but not equal) to the self-similar model, used in the construction of the profiles.
Overall, the scaling relations obtained are in good agreement with the literature \citep{Lovisari2015AA...573A.118L,Mantz2016MNRAS.456.4020M,Schellenberger2017MNRAS.469.3738S,Adami2018A&A...620A...5A,2019ApJ...871...50B,Lovisari2020ApJ...892..102L,Sereno2020MNRAS.492.4528S,Umetsu2020}. 
To obtain XMM-like temperatures and carry out these comparisons, we correct temperatures from the literature following \citet{Schellenberger2015AA...575A..30S}. 
We also convert literature luminosities into the 0.5-2~keV band using the APEC model. 
Additionally, the intrinsic scatter around each model scaling relation is in good agreement with measurements from the literature \citep{Lovisari2015AA...573A.118L,2019ApJ...871...50B,Lovisari2020ApJ...892..102L}. 
We also compare our results to those from alternative numerical methods to mock clusters \citep[Sh10, Za18][]{Shaw2010ApJ...725.1452S, Zandanel2018MNRAS.480..987Z}. 
The methods are overall in good agreement, discrepancies are detailed below. 

We discuss in detail the results for the mass-luminosity (\S~\ref{sec:mlr}), the mass-temperature (\S~\ref{sec:mtr}) and the luminosity-temperature (\S~\ref{sec:ltr}) relations.

\subsubsection{Mass-luminosity relation}
\label{sec:mlr}

\begin{figure}
\centering
\includegraphics[width=.95\columnwidth]{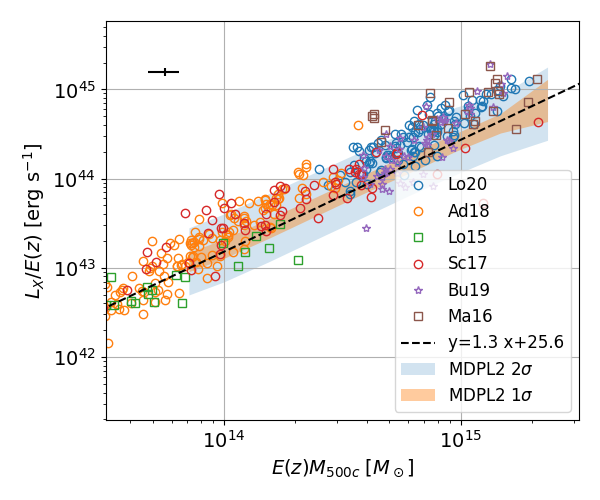}
\includegraphics[width=.95\columnwidth]{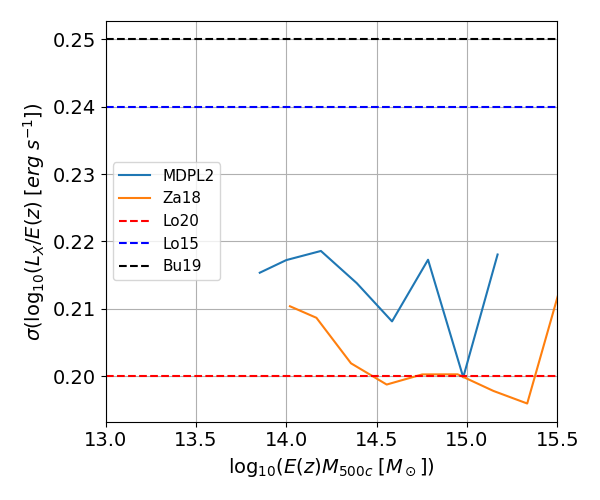}
\includegraphics[width=.95\columnwidth]{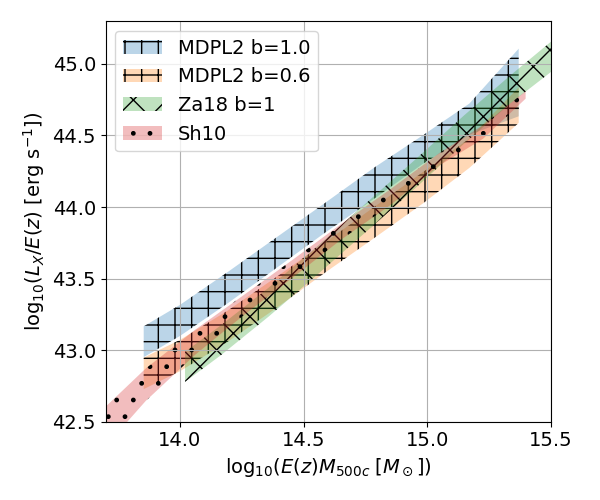}
\caption{\label{fig:cluster:scaling:relation:LX:M500c}
\textbf{Top.} Model scaling relation between X-ray luminosity and mass (shaded areas MDPL2 $1\sigma$, $2\sigma$) compared to existing data sets (open symbols) \citep[Lo15, Ma16, Sc17, Ad18, Bu18, Lo20, Se20 respectively stand for][]{Lovisari2015AA...573A.118L,Mantz2016MNRAS.456.4020M,Schellenberger2017MNRAS.469.3738S,Adami2018A&A...620A...5A,2019ApJ...871...50B,Lovisari2020ApJ...892..102L,Sereno2020MNRAS.492.4528S}. 
The cross on the top-left represent the typical uncertainty on the observed data. 
The agreement is good. 
\textbf{Middle.} 
Intrinsic scatter of the model scaling relation as a function of mass. The scatter is around 0.21. 
This is in excellent agreement with previous studies (0.2, 0.24, 0.25).  
\textbf{Bottom.} 
Model scaling relations between X-ray luminosity and mass for two (extreme) values of the $b_M$ parameter on MDPL2, compared to those from models by Sh10, Za18.}
\end{figure}
 
In the self-similar model, the luminosity is proportional to the mass as $L_X \propto E^2(z) M_{500c} ^{4/3}$. 
The scaling relation obtained here is close, but not consistent with, the self similar model as it has a slope of $1.253 \pm 0.031$. 
This is a consequence of the profile construction methodology. 
Fig. \ref{fig:cluster:scaling:relation:LX:M500c} top panel shows the scaling relation between mass and luminosity of the MDPL2 light-cone. 
We compare with the distribution of measurements \citep{Lovisari2015AA...573A.118L,Mantz2016MNRAS.456.4020M,Schellenberger2017MNRAS.469.3738S,Adami2018A&A...620A...5A,2019ApJ...871...50B,Lovisari2020ApJ...892..102L,Sereno2020MNRAS.492.4528S} and find a good agreement. 
It seems the model relation has a somewhat lower slope and normalization that could be suggested by the data. 
For XXL, the uncertainty on the mass in the data is of order of 15\% and on the X-ray luminosity of order of 10\% (represented by the black cross in top panel of Fig. \ref{fig:cluster:scaling:relation:LX:M500c}). All other surveys have similar uncertainty on the mass and brighter clusters have smaller uncertainties on the X-ray luminosity (a few percents for the brightest).  
Each survey is subject to a different selection function (flux limit, redshift reach, Malmquist bias etc), so the density of data points is not necessarily an accurate indicator of the location of the scaling relation in the figure. 
The model cluster sample being complete, we expect the model scaling relation to be lower than the data points. 
Indeed, the Malmquist bias results in an overestimation of average observed luminosity for given mass. 

We predict that the intrinsic scatter around the mean $\log_{10}(L_X)$ follows a normal distribution with $\sigma=0.21$ with no significant mass dependence, as shown in the middle panel of Fig. \ref{fig:cluster:scaling:relation:LX:M500c}. 
In the literature the intrinsic scatter has been measured at 0.24, 0.25, 0.2 \citep{Lovisari2015AA...573A.118L,2019ApJ...871...50B,Lovisari2020ApJ...892..102L}. 

Mean model scaling relations from Sh10, Za18 are compared to \textsc{MDPL2} for two  values of the $b_M$: 0.6 and 1.0, see Fig. \ref{fig:cluster:scaling:relation:LX:M500c}, bottom panel. 
They are in good agreement. 
The slope we obtain is shallower than that of Za18. 

\subsubsection{Mass-temperature relation}\label{sec:mtr}

\begin{figure}
\centering
\includegraphics[width=.95\columnwidth]{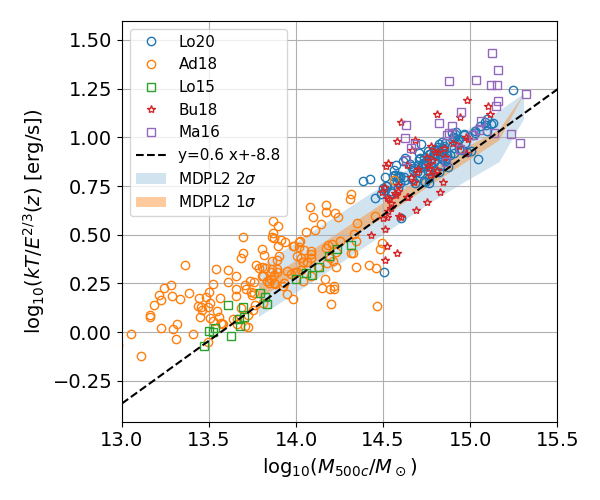}
\includegraphics[width=.95\columnwidth]{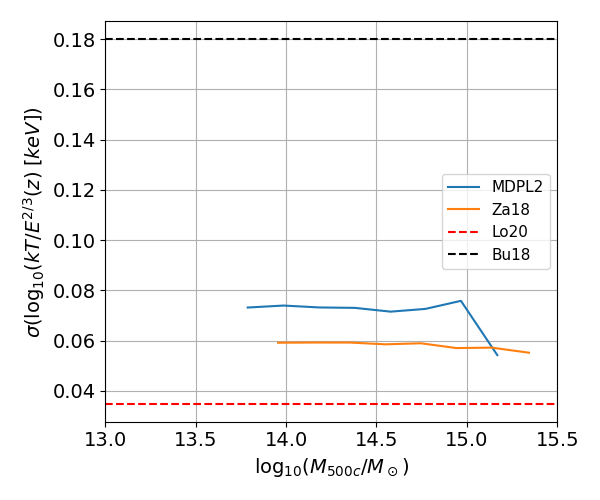}
\includegraphics[width=.95\columnwidth]{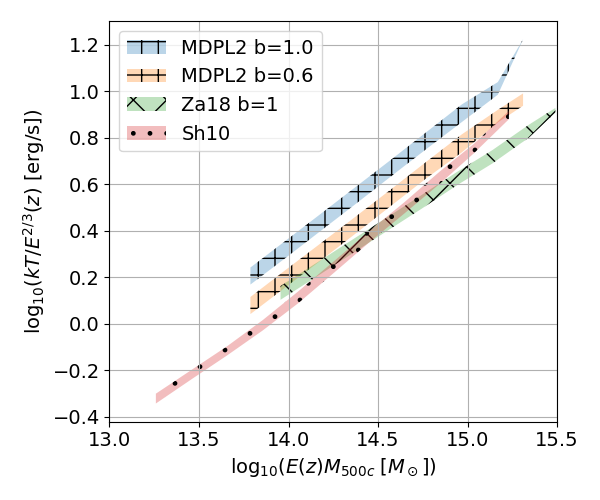}
\caption{\label{fig:cluster:scaling:relation:kT:M500c}
Same as Fig. \ref{fig:cluster:scaling:relation:LX:M500c}. 
\textbf{Top.} Model scaling relation between temperature and mass, the agreement with observations is good. 
\textbf{Middle.} Instrinsic scatter as a function of mass. The mean value of 0.07 is in the middle of values obtained in previous studies: 0.035 and 0.18. 
\textbf{Bottom.} Comparison of model scaling relations.}
\end{figure}

In the self-similar model, the temperature scales with mass as $kT \propto E^{2/3}(z) M_{500c} ^{2/3}$. 
With a slope of $0.644\pm 0.026$, we obtain a scaling relation close to the self similar expectation.
Fig. \ref{fig:cluster:scaling:relation:kT:M500c} shows the scaling relation between mass and temperature. 
As for the previous scaling relations, we compare with the distribution of measurements in the literature and generally find a good agreement. 
The difference with the Sh10 model lies in the different non-thermal pressure fraction. 
The Sh10 model was calibrated on density profiles and the gas mass -- mass relation. 
The non-thermal pressure fraction was taken from simulations \citep{Nelson2014ApJ...792...25N}, which is found to be relatively high compared to the X-COP measurements \citep[e.g.][]{Eckert2019}. 
A change of the non-thermal pressure fraction in the Sh10 model shifts its scaling relations towards that obtained with the method described in this article. 

The predicted intrinsic scatter on the $\log_{10}(kT)$ follows a normal distribution with $\sigma=0.07$. 
In the literature the intrinsic scatter has been measured at 0.035, 0.18 \citep{2019ApJ...871...50B,Lovisari2020ApJ...892..102L} .

\subsubsection{Luminosity-temperature relation}\label{sec:ltr}

\begin{figure}
\centering
\includegraphics[width=.95\columnwidth]{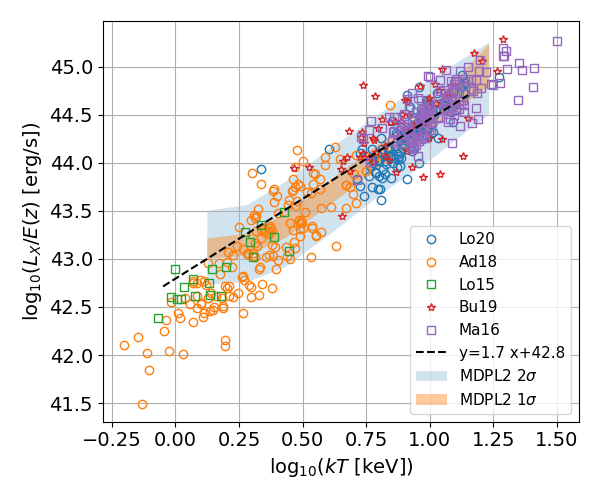}
\includegraphics[width=.95\columnwidth]{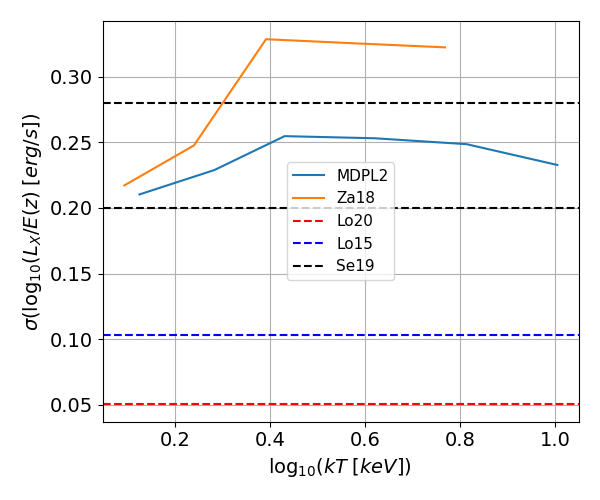}
\includegraphics[width=.95\columnwidth]{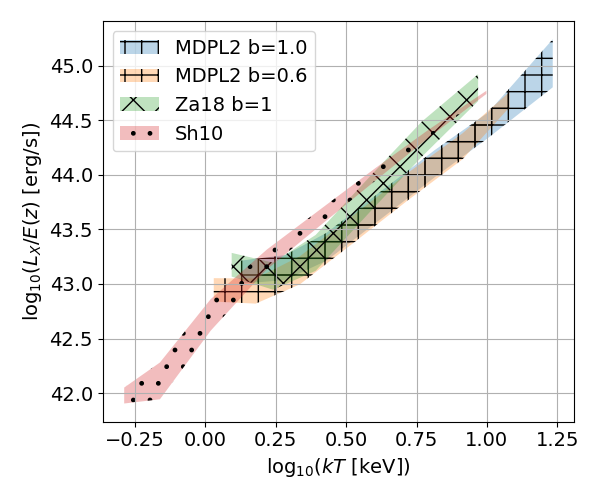}
\caption{\label{fig:cluster:scaling:relation:LX:kT}
Same as Fig. \ref{fig:cluster:scaling:relation:LX:M500c}. 
\textbf{Top.} Model scaling relation between X-ray luminosity and temperature, the agreement with observations is good. 
\textbf{Middle.} Predicted instrinsic scatter as a function of temperature. It is higher than measurements.  
\textbf{Bottom.} Comparison of model scaling relations.}
\end{figure}
 
In the self-similar model, the luminosity scales with temperature as $L_X \propto E^{2/3}(z) (kT) ^{2}$. 
With a slope of $1.659\pm0.098$, we obtain a scaling relation close to (but not  consistent with) that. 
By construction, the scaling relation predicted is close to that of \citet{Giles2016,Lieu2016AA...592A...4L}, based on the XXL data.
Fig. \ref{fig:cluster:scaling:relation:LX:kT} shows the scaling relation between temperature and luminosity. 
We compare with the distribution of measurements from \citep{Lovisari2015AA...573A.118L,Mantz2016MNRAS.456.4020M,Adami2018A&A...620A...5A,2019ApJ...871...50B,Lovisari2020ApJ...892..102L} and find a good agreement. 
The turn over of the relation at low temperature is due to the Malmquist bias in the XXL data used, future inclusion of lower mass objects in the covariance matrix will enable to correct this effect and extend the model to lower masses.

The predicted intrinsic scatter on the $\log_{10}(L_X)$ follows a normal distribution with $\sigma=0.25$. 
In the literature the intrinsic scatter has been measured at 0.2 \citep{Lovisari2015AA...573A.118L} and 0.2-0.28 \citep{Sereno2019AA...632A..54S}.

\subsection{Number density}
\label{subsection:number:density}
\begin{figure}
\centering
\includegraphics[width=.9\columnwidth]{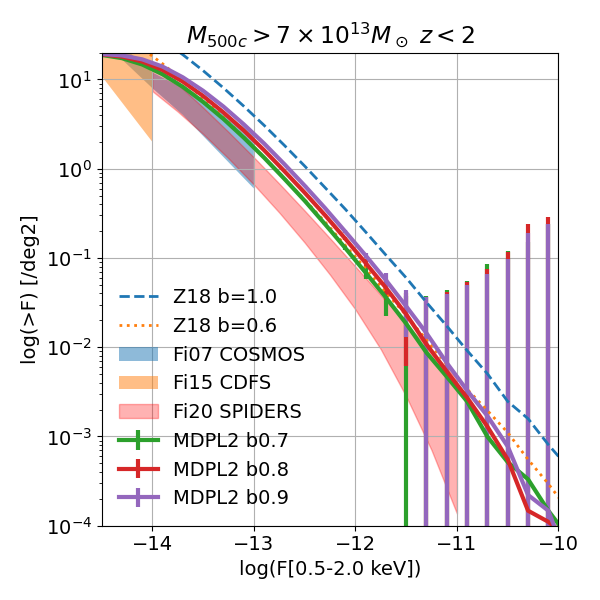}
\caption{\label{fig:cluster:logNlogS}
Number density per square degree of clusters with a flux greater than $F$ as a function of flux $F$. 
We show the curve corresponding to the \textsc{MDPL2} model clusters. 
These model clusters extend up to redshift 2 and are by construction limited to $M_{500c}>7 \times 10^{13}M_\odot$. 
We compare with measurements from \citet{Finoguenov2007,Finoguenov2015,Finoguenov2019}. 
The agreement is good. 
}
\end{figure}

The number density of clusters (per square degree, often named $\mathrm{\log\; N \; - \log \; S}$) as a function of X-ray flux (soft band) for clusters with $M_{500c}>7 \times 10^{13}M_\odot$ is shown on Fig. \ref{fig:cluster:logNlogS}. 
We find the prediction to be in agreement with current measurements \citep{Finoguenov2007,Finoguenov2015,Finoguenov2019,Bohringer2017}. 

The cosmological parameter used in the N-body simulations (combination of $\Omega_m\sim 0.31$ and $\sigma_8\sim 0.8228$) are known to produce `too many' clusters \citep[\textit{e.g.}][]{Zandanel2018MNRAS.480..987Z}, we indeed see that the predicted $\mathrm{\log\; N \; - \log \; S}$ lies at the upper boundary of observations. 
The $b_M$ parameter has quite an impact here. 
The higher the $b_M$ the more bright clusters will be present in the mock, thus their density will increase. 

\subsection{Simulated \emph{eROSITA} photons} 
\label{subsec:fullsky:photons}

We use the \textsc{sixte} software package for X-ray telescope observation simulations
\citep{Dauser2019AA...630A..66D}\footnote{\url{http://www.sternwarte.uni-erlangen.de/research/sixte/}}.
It produces simulated event files for mission studies.
We create a full sky event list using the observational strategy from \emph{eROSITA}. 
With a healpix pixelisation nested scheme, we split the sky into 768 equal area pixels of approximately 53 deg$^2$ each \citep{Gorski2005ApJ...622..759G}. 
The co-latitude is $90 - {\rm Dec.}$ and the longitude is R.A.

Along with clusters, we model a population of AGN. 
We release the set of AGN and cluster photons. 
The AGN model is described in \citep{Comparat2019MNRAS.tmp.1335C}. 
A notable difference in the AGN model with respect to its previous incarnation is that they are also assigned to dark matter sub-halos. 
A caveat to note is that the incidence of AGN within clusters or groups (its redshift and mass dependence) is relatively poorly known \citep{Martini2013ApJ...768....1M,Koulouridis2018AA...620A..20K}. 
In this incarnation, of the models, there is no cross-talk between the cluster and the AGN model. 
Future developments of the AGN model will include a set of scenarii of AGN-cluster co-evolution. 

\begin{figure*}
\centering
\includegraphics[width=1.9\columnwidth]{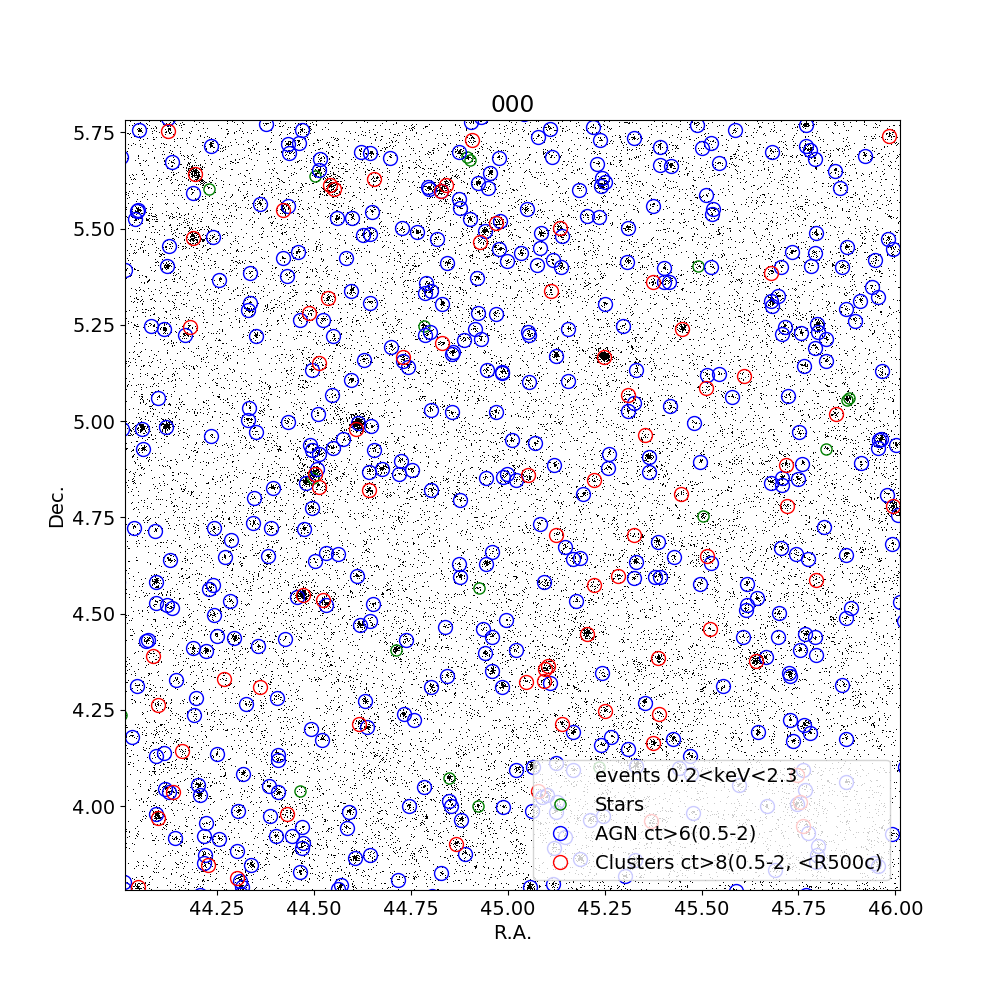}

\caption{\label{fig:events:sky000}
Declination vs. Right Ascension of the photons observed (black dots) in the 0.2-2.3~keV band after 4 years of \emph{eROSITA} observations on the 2x2 degrees center of the field 0 (out of 768). 
A fraction of the sources present in the catalogue are identified by circles. 
It shows the clusters that emmitted more than 8 counts (red), the AGN that emmitted more than 6 counts (blue), and the stars (green).}
\end{figure*}

\begin{figure*}
\centering

\includegraphics[width=1.9\columnwidth]{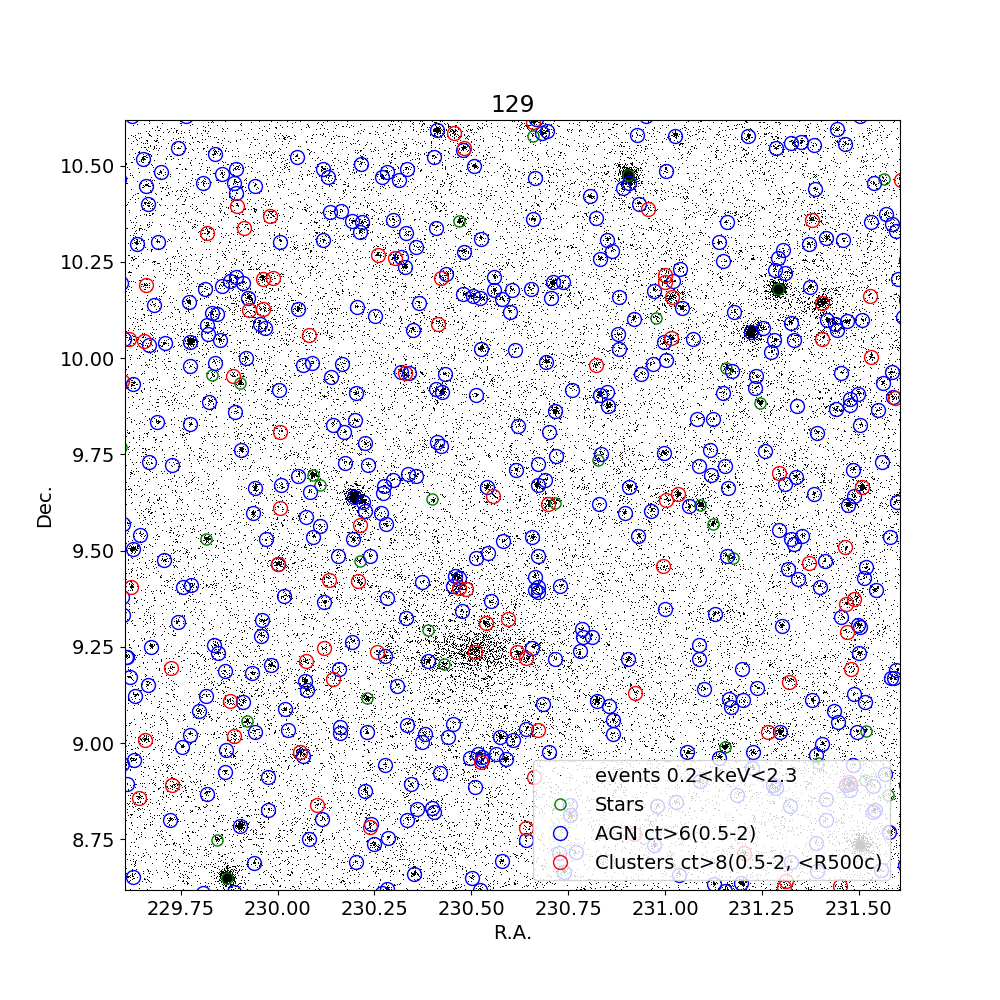}
\caption{\label{fig:events:sky129}
Same as Fig. \ref{fig:events:sky000} for field 129.}
\end{figure*}

Figs. \ref{fig:events:sky000} and \ref{fig:events:sky129} show the events related to the clusters and AGNs on a couple $2^\circ \times 2^\circ$ fields. 
The diversity of the cluster population is well represented. 
Understanding the co-evolution of AGN and clusters seems to be key to turn these event lists back into unbiased catalogues of the large scale structure. 

Ongoing \emph{eROSITA} studies detailing the procedure of detection and catalogue creation intensively use these simulations to understand the trade-off between purity and completeness. 
These also enable an in-depth discussion of the line-of-sight confusion between the various sources as a function of exposure time.

To enable the study of systematic errors throughout the data flow, we aggregate photon statistics (number of photons in a set of apertures emitted by the source itself or other nearby sources) in the simulated catalogues. 
 
\section{Summary and outlook}
\label{sec:summary}

We presented a novel empirical approach to model X-ray profiles of cluster and group-size halos. 
We used the MultiDark \citep[\textsc{HMDPL, MDPL2, SMDPL}][]{Klypin2016} and the UNIT \citep[\textsc{UNIT1, UNIT1i}][]{Chuang2019} dark matter only simulations to create light-cones spanning the full sky. 
We have created a new method to simulate the X-ray light emitted by the hot gas in galaxy clusters. 
This model computes the emissivity and image for the clusters in the light-cones. 

We find that the mean scatter around the profile as a function of scale is in agreement with the latest measurements from \citet{Ghirardini2019}. 
The model scaling relations, their scatter and the number density of clusters are in good agreement with observations and other models. 
We obtain a scatter of 0.21 (0.07, 0.25) for the X-ray luminosity -- mass (temperature -- mass, temperature -- luminosity) model scaling relations.
Finally, using the AGN model from \citet{Comparat2019MNRAS.tmp.1335C}, we predict the full sky distribution of photons to be observed by \emph{eROSITA}. 

This is the first building block of an end to end validation of the \emph{eROSITA} work flow towards measuring the large scale structure traced by X-rays. 
This will enable accurate estimations of various systematic uncertainties associated with cluster astrophysics (e.g., cool-cores, non-thermal pressure, gas clumping) and those due to observation and analysis methods (e.g., projection effects). 
This model data presented here will support upcoming studies of the \emph{eROSITA} source detection process and selection function computation.

We will extend the method to lower mass halos by extracting X-ray profiles of groups from the deep Chandra on CDFS \citep{Finoguenov2015} and COSMOS \citep{Gozaliasl2019}. 
We will extend the discussion on scaling relations to core-excised quantities and count rates in future studies. 
Towards a more quantitative validation of the method, we will apply it to the dark matter haloes of hydro-dynamical simulations and compare its predictions with that of the hydro-dynamical simulation \citep[e.g.][]{biffi_2018MNRAS.tmp.2317B}.

In the future, we plan to create models of foregrounds and backgrounds, relative to the cluster and AGN population, to best understand their selection function. 
We plan to investigate how to mock low redshift galaxies, stars, the large and small magellanic clouds and the diffuse emission of the Milky Way.

The combined X-ray catalogues of AGN and galaxy clusters and groups produced from our model will also open up new avenues in the study of co-evolution between AGN and galaxy clusters and groups with \emph{eROSITA} and other future X-ray surveys by cross-correlating AGN and cluster signals. This will lead to better understanding of the role of AGN feedback in physics of intra-cluster and intra-group medium. 
The modelling pipeline presented in the current paper can also applied to other wavelengths, such as the microwave in modeling the Compton-y signals from galaxy clusters and groups, and the modeling of gravitational lensing signals, using the same N-body lightcone. 
This will be useful and essential for generating predictions for multiwavelength observations for constraining outstanding issues in cluster astrophysics, such as non-thermal pressure and hydrostatic mass bias, and clumping. 
This will help \emph{eROSITA} to maximize its scientific returns through synergies with other ongoing microwave and optical surveys.

\section*{Acknowledgements}

DN acknowledges the Max-Planck-Institut f\"{u}r Astrophysik for hospitality when this work was initiated.

ET was supported by ETAg grants IUT40-2, PRG1006 and by EU through the ERDF CoE TK133. 

GY acknowledges financial support by  MICIU/FEDER under research  grant PGC2018-094975-C21. 

The authors gratefully acknowledge the Gauss Centre for Supercomputing and the Partnership for Advanced Supercomputing in Europe (PRACE) for funding the MultiDark simulation project by providing computing time on the GCS Supercomputer SuperMUC at Leibniz Supercomputing Centre, Germany.

The UNIT simulations were  run at the MareNostrum Supercomputer hosted by the  Barcelona Supercomputing Center (BSC), Spain,  thanks to the  PRACE  project  grant  number 2016163937.

The CosmoSim database is a service of the Leibniz-Institute for Astrophysics Potsdam (AIP).
The MultiDark database was developed in cooperation with the Spanish MultiDark Consolider Project CSD2009-00064.

This project made use of 
\textsc{gawk}\footnote{{https://www.gnu.org/software/gawk/}}, 
\textsc{python}3\footnote{\url{https://www.python.org}}, 
\textsc{astropy} \citep{Astropy2013AA, Astropy2018AJ}, 
\textsc{topcat/stilts}\footnote{\url{http://www.star.bris.ac.uk/~mbt/stilts/}} \citet{2006ASPC..351..666T}, 
\textsc{slurm}\footnote{\url{https://slurm.schedmd.com/publications.html}}, 

This research has made use of the SIXTE software package \citep{Dauser2019AA...630A..66D} provided by ECAP/Remeis observatory (\url{https://github.com/thdauser/sixte}).
%%%%%%%%%%%%%%%%%%%%%%%%%%%%%%%%%%%%%%%%%%%%%%%%%%
%%%%%%%%%%%%%%%%%%%% REFERENCES %%%%%%%%%%%%%%%%%%
\bibliographystyle{mnras}
\bibliography{00_main}

%%%%%%%%%%%%%%%%%%%%%%%%%%%%%%%%%%%%%%%%%%%%%%%%%%
%%%%%%%%%%%%%%%%% APPENDICES %%%%%%%%%%%%%%%%%%%%%
 
\appendix

\section{AGN model}

In this work we upgrade the model from \citet{Comparat2019MNRAS.tmp.1335C}. 
We use the \citet[\textsc{Universe Machine}][]{Behroozi2019MNRAS.488.3143B} empirical galaxy model to compute and attach the galaxy quantities, such as stellar mass, star formation rate, for each dark matter halo. 
Note that there exists more galaxy models run on the \textsc{MDPL2} simulation \citep{Stoppacher2019MNRAS.486.1316S}. 
Each model has advantages and disadvantages. 
The main reason for choosing the \textsc{Universe Machine} is that this model is constrained to reproduce the fraction of quenched galaxies as a function of stellar mass, which is an important feature for future studies of galaxies in clusters.

For AGNs, we follow the works by \citet{2018MNRAS.tmp.3272G,Comparat2019MNRAS.tmp.1335C} without a prior on the fraction of sub-halos (i.e. we treat all sub-halos in the simulation as if they were haloes). 
We then consider sub-halos as possible AGN hosts in the simulation, when computing the duty cycle. As the periodic boundary condition is respected (while replicating the simulation boxes) these mocks are suited for clustering studies on the full sky.

\section{data and software}

%\subsection{Access}

The light-cones and event files produced are made available in two locations: MPE/MPG, Germany\footnote{\url{http://www.mpe.mpg.de/~comparat/eROSITA_mock/}} and skies and universes, Granada, Spain\footnote{\url{http://skiesanduniverses.org/}} \citep{skies_universes_2017arXiv171101453K}. 
We make available the catalogues and event lists produced based on the \textsc{MDPL2} simulation. 
The largest part of the pipeline developed is public and divided in two repository. The first to create the light cones and the mock catalogues
\footnote{\url{https://github.com/ygolomsochtiwsretsulc/mocks_high_fidelity}} and the second to draw cluster model profiles 
\footnote{\url{https://github.com/domeckert/cluster-brightness-profiles}}. 
 
\end{document}